\documentclass{article}

\usepackage[utf8]{inputenc}
\usepackage{14lomcon}
\usepackage{cite}
\usepackage{epsfig}
\usepackage{psfig}
\usepackage{amsmath}

\begin{document}

\title{HIGH-ENERGY NEUTRINOS FROM GALACTIC SOURCES}

\author{ Alexander Kappes \footnote{e-mail: kappes@physik.uni-erlangen.de}}

\address{Friedrich-Alexander-Universität Erlangen-Nürnberg, Erlangen
  Centre for Astroparticle Physics, Erwin-Rommel-Str. 1, 91058
  Erlangen, Germany}


\maketitle\abstracts{ Even 100 years after the discovery of cosmic
  rays their origin remains a mystery. In recent years, TeV gamma-ray
  detectors have discovered and investigated many Galactic sources
  where particles are accelerated up to energies of 100 TeV.  However,
  it has not been possible up to now to identify these sites
  unambiguously as sources of hadronic acceleration. The observation
  of cosmic high-energy neutrinos from these or other sources will be
  a smoking-gun evidence for the sites of the acceleration of cosmic
  rays.}
\section{Introduction}
Up to now, information on objects in our galaxy and beyond
has nearly exclusively been obtained using electromagnetic waves as
cosmic messengers. Modern astronomy started in 1610 with optical
photons, when Galileo Galilei took one of the first telescopes and
pointed into the sky. Only in the last century, people started to
extended the observations to lower and higher energies of the
electromagnetic spectrum, where today's instruments span an enormous
range of 20 orders of magnitude from radio waves to TeV gamma-rays.

In addition, we know from measurements of the cosmic-ray spectrum that
there exist sources in the universe which accelerate protons or
heavier nuclei up to energies of $\sim 10^{20}$\,eV, $10^7$ times higher
than the most energetic man-made accelerator, the LHC at CERN. These
highest energies are believed to be reached in extra-Galactic sources
like gamma-ray bursts or active galactic nuclei whereas Galactic
sources like supernova remnants (SNRs) or micro-quasars are thought to
accelerate particles at least up to energies of $3\times10^{15}$\,eV,
also called the ``knee'' region of the cosmic-ray spectrum. Though
these phenomena are quite different, the basic acceleration mechanism
is believed to be very similar. Particles are injected into shock
fronts which develop when fast moving matter collides with other
matter. In these shock fronts, the injected particles are accelerated
in a repeating process (Fermi acceleration) where the energy gain per
cycle is only small.  Other phenomena where a ``one-shot''
acceleration of cosmic rays to high energies might take place are
objects with strong magnetic fields (up to $10^{15}$\,G) like pulsars
and magnetars. However, despite the detailed measurements of the
cosmic-ray spectrum and 100 years after their discovery by Victor
Hess, we still do not know what the sources of the cosmic rays are as
they are deflected in the Galactic and extra-Galactic magnetic fields
and hence have lost all information about their origin when reaching
Earth. Only at the highest energies beyond $\sim 10^{19.6}$\,GeV
cosmic rays might retain enough directional information to locate
their sources.

Alternative messengers for locating the sources of the cosmic rays
must have two distinct properties: they have to be electrically
neutral and essentially stable. Only two of the known elementary
particles meet these requirements: photons and neutrinos. Both
particles are inevitably produced when the accelerated protons or
nuclei collide with matter or photons inside or near the source. In
these reactions neutral and charged pions are produced which then
decay into high-energy photons and neutrinos, respectively. The well
known ratio between the production of neutral and charged pions yields
a direct link between the photon and neutrino flux from proton-proton and
proton-gamma interactions, respectively.

The number of sources detected in TeV gamma-rays has increased
dramatically during the last decade which was made possible by
significant advances in the technique of air Cherenkov telescopes.
Today, over 100 sources are known, both Galactic (e.g., SNRs,
micro-quasars, pulsars) and extra-Galactic (e.g., active galactic
nuclei, starburst galaxies). However, it has not been possible up to now to
unambiguously identify these sources as sites of cosmic-ray
acceleration as the observed TeV photons could also have been
generated in the up-scattering of photons in reactions with
accelerated electrons (inverse-Compton scattering). High-energy
neutrinos, on the other hand, are only produced in reactions of
accelerated hadrons and hence are a smoking-gun evidence for the
sources of cosmic rays.

\section{Potential Galactic neutrino sources}
From the discussion above it follows naturally that TeV gamma-ray
sources are prime candidates for high-energy neutrino emission.
Today, the list of Galactic objects with observed TeV gamma-ray
emission contains a.o. SNRs, pulsar wind nebulae (PWNe), binary
systems and molecular clouds. In the following, we will discuss the
prospects for neutrino emission from these objects and the expected
fluxes. For an overview and more details about the sources see
\cite{rpp:71:096901} and references therein.

\begin{figure}[t]
\includegraphics[width=6cm]{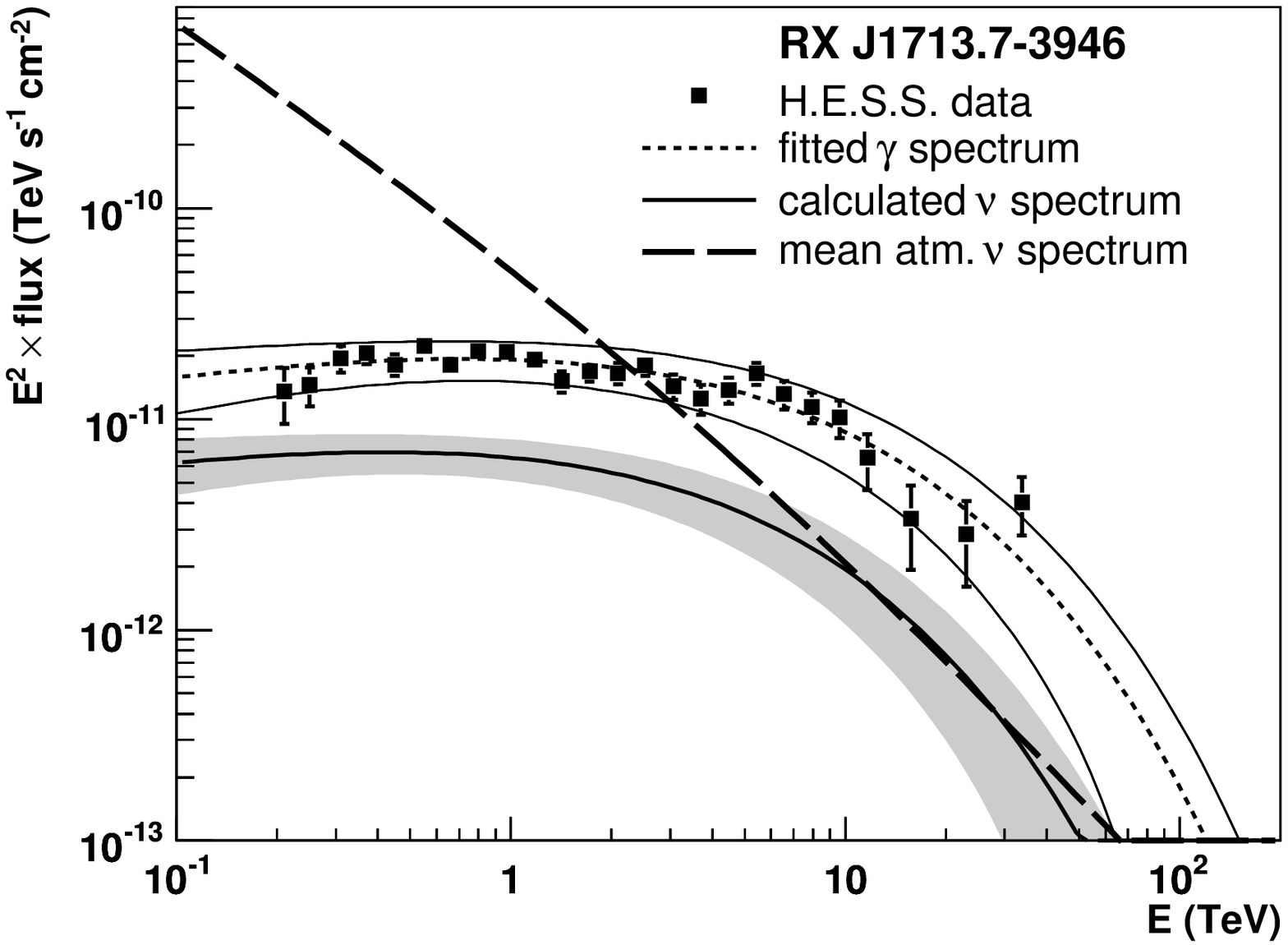}
\includegraphics[width=4.8cm]{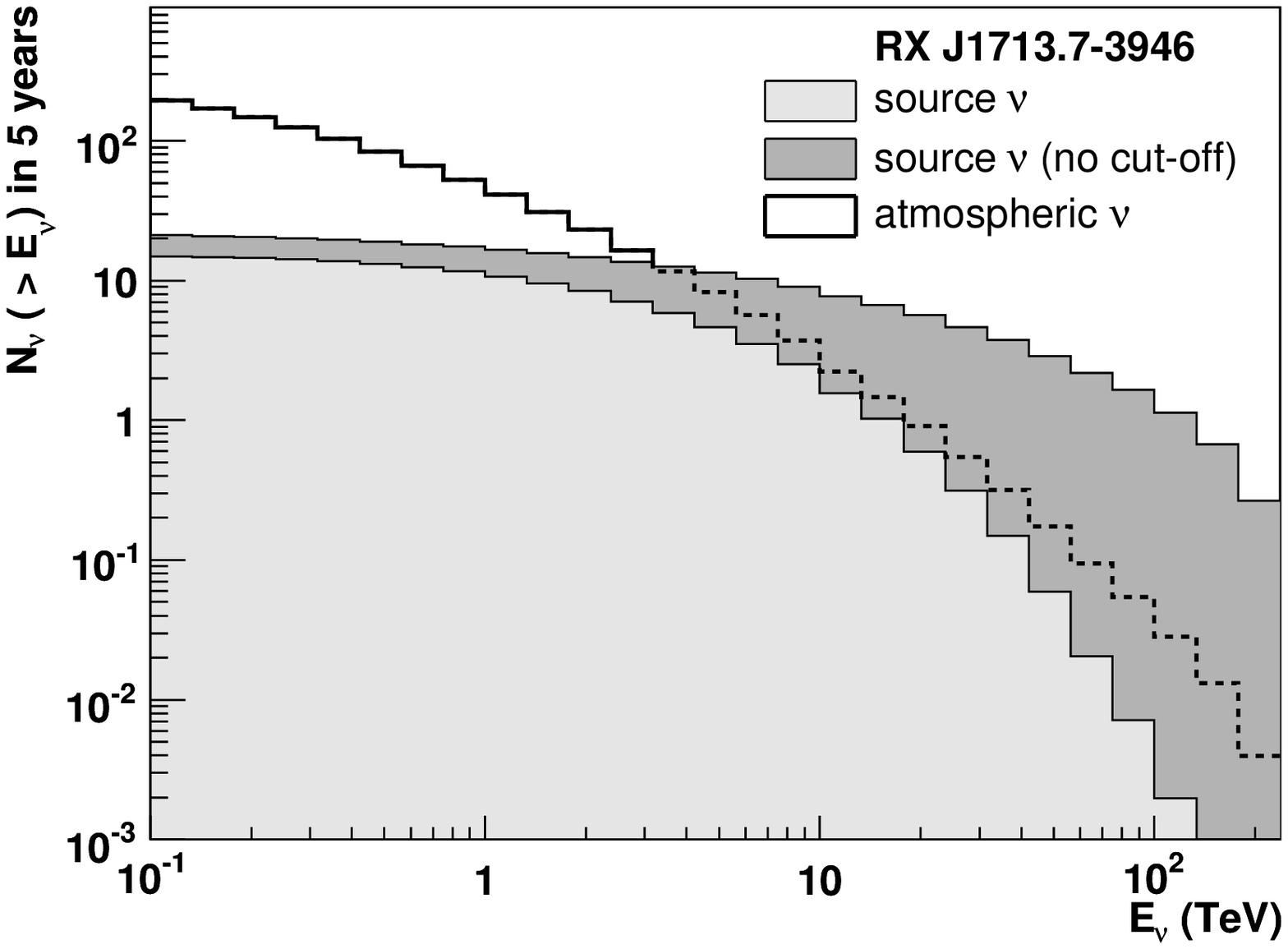}
\caption{Left: Measured photon and expected neutrino flux from
  RX\,J1713.7$-$3946. The area formed by the solid lines and the
  shaded area represent the uncertainty (including systematics) on the
  respective spectrum. Right: Expected event rates from
  RX\,J1713.7$-$3946 in a cubic-kilometer sized neutrino telescope
  with and without a high-energy cut-off. Also shown is the rate of
  atmospheric neutrinos. Taken from \cite{apj:656:870}.}
\label{rxj1713}
\end{figure}

\vspace{1mm}{\bf Supernova remnants:} SNRs are the prime candidate for
the acceleration of the Galactic cosmic rays. This has several
reasons: There exists a plausible model for the acceleration of
hadrons in the shock fronts generated when the ejecta of the supernova
interact with the surrounding medium. Furthermore, energetic
considerations show \cite{jpcs:171:12014} that the observed supernova
rate of about 3 per century and a conversion of 10\% of the
$10^{51}$\,erg released per nova into cosmic-ray acceleration matches
quite well the energy required to keep the energy in cosmic rays
constant over time.  Up to now, 11 SNRs have been observed in TeV
gamma-rays.  Under the assumption that all gamma-ray photons origin
from the interaction of accelerated hadrons and utilizing the
connection between gamma-rays and neutrinos one can make rather
precise predictions for the expected neutrino flux from these sources
\cite{apj:656:870}.  An example of such a calculation is shown in the
left plot of Fig.~\ref{rxj1713}.  Using the effective muon neutrino
area of a neutrino telescope one can calculate the integrated number
of observed events as a function of the threshold energy.
This is displayed in the right plot of Fig.\,\ref{rxj1713} for a water
Cherenkov neutrino telescope with an instrumented volume of 1\,km$^3$.
The importance of the cut-off at high energies for the detectability
of the source becomes apparent when comparing the light and dark
shaded areas with the expected background from atmospheric neutrinos.
The strong contribution of high-energy neutrinos to the number of
detected events origins from the strong rise of the effective area
with the neutrino energy which is mainly caused by the increasing
neutrino-nucleon cross section and the extended range of the muons
produced. On the other hand, the number of background events scales
with the size of the emission region which in the case of
RX\,J1713.7$-$3946 is quite large with $1.3^\circ$ diameter. These two
features, the low energy cut-off and the large emission region, make
the detection of this and similar sources like RX\,J0852.0$-$4622 with
neutrino telescopes challenging. In addition, there exits a population
of SNRs where no cut-off could be determined from the measured
gamma-ray spectrum due to the large error bars at high energies.
However, these spectra are rather steep and only a small number of
events comparable to those from atmospheric neutrinos is expected
\cite{apj:656:870}.

\vspace{1mm}{\bf Binary systems:} The binary systems observed in
gamma-rays consist of a compact object (either a neutron star or a
black hole) and a companion star. The compact object accretes material
from the companion star producing jets where particles are
accelerated. One of these objects is LS\,5039 discovered with the
H.E.S.S.\ telescope. The system has a period of 3.9\,days and the TeV
gamma-ray emission region is point-like for neutrino telescopes. A
calculation of the expected neutrino event rate from the measured
gamma-ray flux yields less than one event per year for a
cubic-kilometer neutrino telescope \cite{apj:656:870}.  However, due
to the dense photon fields originating from the companion star the
measured gamma-ray flux may be significantly suppressed by up to a
factor 100 \cite{jpcs:39:408}. If this is indeed true then this and
similar sources would be one of the most promising targets for
neutrino telescopes.

\vspace{1mm}{\bf Pulsar wind nebulae:} A PWN is powered by the wind
from a pulsar which streams into the ambient medium creating shock
fronts. In these shock fronts particles are accelerated. In general,
PWNe are believed to accelerate mainly electrons. In \cite{aa:451:l51}
however, the authors argue that there also might be a significant
fraction of nuclei in the pulsar wind. In this case PWNe like Vela X
pose interesting targets for neutrino telescopes.

\vspace{1mm}{\bf Molecular clouds:} Gamma-ray emission from molecular
clouds origins from the interaction of cosmic rays from a nearby
source, e.g., a SNR, with the nuclei in the cloud. Hence, molecular
clouds are a ``guaranteed'' source of neutrino emission. Such an
emission region was located by H.E.S.S.\ near the Galactic center.
Unfortunately, the expected neutrino fluxes derived from the measured
gamma rays are rather low and the emission region is quite large
making the detection of this and other molecular clouds in neutrinos
very challenging.

\section{The missing Pevatrons}
The existence of the ``knee'' in the cosmic ray spectrum tells us that
there must exist Galactic cosmic-ray sources which accelerate protons
up to energies of several PeV. These ``Pevatrons'' will produce pionic
gamma rays whose spectrum extends to several hundred TeV without
cut-off in interactions with the interstellar medium. However, none of
the observed gamma-ray sources shows such a spectrum. One reason for
this observation could be the fact that the highest energies in SNRs
are only reached during the first few hundred years after the
supernova explosion (a SNR has a typical lifetime of the order of
10,000 years). In this case, it could simply be possible that
currently there does not exist an observable SNR in this early phase.
Nevertheless, the detection of these cosmic rays could still be
possible by observing the gamma rays produced in their interaction
with interstellar medium, in particular, with dense molecular clouds
as shown in \cite{apjl:665:l131}. In \cite{pr:d78:063004} the authors
argue that some of the gamma-ray sources discovered by Milagro
\cite{apj:664:L91} might origin from such secondary interactions of
cosmic rays from Pevatrons. Future measurements with air Cherenkov and
neutrino telescopes have to show whether this is true. On the other
hand, one of the suggested sources, MGRO\,1908+06, has now been
plausibly associated with a pulsar \cite{apj:711:64}, disfavoring it
as a region of proton acceleration.

\section{Sensitivity of current and future neutrino telescopes}
All current event rate calculations indicate that neutrino telescopes
with instrumented volumes of cubic kilometer size are necessary to
identify the first cosmic source of high-energy neutrino emission. Due
to the large background of downgoing muons produced in interactions of
cosmic rays with the Earth's atmosphere, neutrino telescopes have
their highest sensitivity when looking through the Earth. Hence, in
order to cover the full sky with high sensitivity both a telescope in
the Northern and in the Southern hemisphere is necessary.

\begin{figure}[t]
\centering
\includegraphics[width=6.cm]{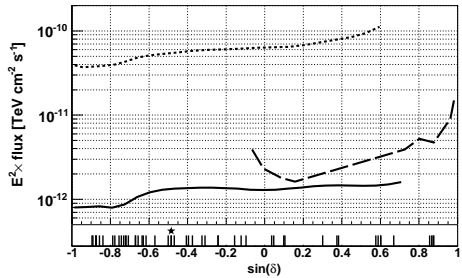}
\caption{Upper part: Sensitivity at 90\% CL of neutrino telescopes to
  an $E^{-2}$ neutrino flux as a function of the source declination:
  ANTARES (dotted, predicted) \cite{arXiv:1002.0754}, IceCube
  (dashed, predicted) \cite{proc:icrc09:dumm:1}, KM3NeT (solid,
  not final), \cite{proc:vlvnt09:katz:1}. Lower part: declination of
  Galactic objects with observed TeV gamma-ray emission. The position
  of the Galactic Center is marked with a star.}
\label{ps_sensitivity}
\end{figure}

Currently, the most sensitive neutrino telescope in the Northern
hemisphere is the ANTARES detector \cite{arXiv:1002.0754}, installed
at a depth of 2500\,m off the coast of south France in the
Mediterranean Sea. The detector instruments a volume of about
0.01\,km$^3$ and therefore is probably too small for cosmic neutrino
detection. In the Southern hemisphere, the currently largest and hence
most sensitive neutrino telescope, IceCube, is close to completion
planned for next year. With an instrumented volume of 1\,km$^3$ it is
the first cubic-kilometer-class detector. The sensitivities of both
experiments to an $E^{-2}$ neutrino flux are shown in
Fig.~\ref{ps_sensitivity}.  It is apparent, that in order to achieve a
high-sensitivity coverage over the full sky a cubic kilometer class
detector in the Northern hemisphere is needed which will cover most of
the Galactic plane including the Galactic center.  Such a detector
with an instrumented volume of about 5\,km$^3$, KM3NeT
\cite{km3net:cdr:2008}, is currently in its planning phase. Data
taking is planned to start around 2014.

\section{Summary and outlook}
The observation of cosmic high-energy neutrinos will open a new window
to the universe and help to solve long-standing mysteries like the
question of the origin of the cosmic rays. With the recent advances in
gamma-ray astronomy we were able to identify many good Galactic
candidate sources of neutrino emission. However, due to the expected
low fluxes and the early high-energy cut-offs of the predicted
neutrino spectra the detection of these sources is challenging.

IceCube will be the first detector to advance into the ``discovery
region'' but will likely only scratch it. Therefore, a factor five to
ten more sensitive detector is needed in the Northern hemisphere which
will also cover the inner Galactic Plane and Center. This detector,
KM3NeT, is supposed to start data taking around 2014. With IceCube and
KM3NeT combined, neutrino astronomy will hopefully soon become
reality.

\section*{Acknowledgments}
Supported by the BMBF under project 05A08WEA. The author also acknowledges the support by
the EU Marie Curie OIF program.

\end{document}